\documentclass[%
 aip,
 amsmath,amssymb,
 reprint,%
]{revtex4-1}

\usepackage{graphicx}
\usepackage{dcolumn}
\usepackage{bm}

\usepackage{xcolor}
\usepackage{upgreek}
\usepackage{hyperref}

\newcommand{\crit}{\mathrm{cr}}
\newcommand{\D}{\mathrm{d}}
\newcommand{\del}{\partial}
\newcommand{\ssb}{\mathbf{s}}
\newcommand{\Eb}{\mathbf{E}}
\newcommand{\Bb}{\mathbf{B}}
\newcommand{\vb}{\mathbf{v}}
\newcommand{\jb}{\mathbf{j}}
\newcommand{\Omegab}{\bm{\Omega}}

\usepackage[utf8]{inputenc}
\usepackage[T1]{fontenc}
\usepackage{mathptmx}
\usepackage{etoolbox}

\makeatletter
\def\@email#1#2{%
 \endgroup
 \patchcmd{\titleblock@produce}
  {\frontmatter@RRAPformat}
  {\frontmatter@RRAPformat{\produce@RRAP{*#1\href{mailto:#2}{#2}}}\frontmatter@RRAPformat}
  {}{}
}%
\makeatother
\begin{document}

\preprint{AIP/123-QED}

\title[Laguerre-Gaussian pulses for spin-polarized ion beam acceleration]{Laguerre-Gaussian pulses for spin-polarized ion beam acceleration}

	\author{Lars Reichwein}
	\email{l.reichwein@fz-juelich.de}
     \affiliation{Peter Gr\"{u}nberg Institut (PGI-6), Forschungszentrum J\"{u}lich, 52425 J\"{u}lich, Germany}
	\affiliation{Institut f\"{u}r Theoretische Physik I, Heinrich-Heine-Universit\"{a}t D\"{u}sseldorf, 40225 D\"{u}sseldorf, Germany}

 \author{Tong-Pu Yu}
 \affiliation{College of Science, National University of Defense Technology, Changsha 410073, People's Republic of China}

	\author{Alexander Pukhov}
	\affiliation{Institut f\"{u}r Theoretische Physik I, Heinrich-Heine-Universit\"{a}t D\"{u}sseldorf, 40225 D\"{u}sseldorf, Germany}

    \author{Markus B\"{u}scher}
	\affiliation{Peter Gr\"{u}nberg Institut (PGI-6), Forschungszentrum J\"{u}lich, 52425 J\"{u}lich, Germany}
	\affiliation{Institut f\"{u}r Laser- und Plasmaphysk, Heinrich-Heine-Universit\"{a}t D\"{u}sseldorf, 40225 D\"{u}sseldorf, Germany}

\date{\today}

\begin{abstract}
     Polarized particle sources have a plethora of applications, ranging from deep-inelastic scattering to nuclear fusion. One crucial challenge in laser-plasma interaction is maintaining the initial polarization of the target. Here, we propose the acceleration of spin-polarized Helium-3 from near-critical density targets using high-intensity Laguerre-Gaussian laser pulses. Three-dimensional particle-in-cell simulations show that Magnetic Vortex Acceleration with these modes yields higher polarization on the $90\%$-level compared to conventional Gaussian laser pulses, while also providing low-divergence beams.
\end{abstract}

\maketitle

\section{Introduction}

    In recent years, plasma-based, spin-polarized particle sources have gained significant interest due to their applications ranging from high-energy \cite{Glashausser1979} to surface physics \cite{Tusche2024}. 
    Moreover, Kulsrud \textit{et al.} showed that polarized reactants in nuclear fusion can increase the corresponding cross-section.\cite{Kulsrud1982} In the case of the reaction $d + t \to \alpha + n$, the cross-section is approx. 1.5 times larger.

    Accordingly, polarized fusion is being investigated by several groups as a possible pathway to higher fusion yields.\cite{Baylor2023, Parisi2024, Heidbrink2024}

    From a purely accelerator-based perspective, Magnetic Vortex Acceleration (MVA) \cite{Jin2020, Reichwein2021, Zou2025} and Collisionless Shock Acceleration (CSA) \cite{Yan2023, Reichwein2024} have been investigated intensively as possible acceleration schemes for polarized protons and other ions.
    A review of the state-of-the-art for this field has been given by Reichwein \textit{et al.}\cite{Reichwein2024review}

    Other schemes commonly used for ion acceleration like Target Normal Sheath Acceleration are currently not available in the context of polarized beams, since the targets have to be pre-polarized using mechanisms like Metastability Exchange Optical Pumping.\cite{Mrozik2011}
    
    Thus, available targets are currently limited to near-critical densities.\cite{Fedorets2022, Sofikitis2025} The first successful plasma-based acceleration of polarized Helium-3 was shown by Zheng \textit{et al.}, where the dominant acceleration mechanism was identified as Coulomb explosion due to the specific target density and the interaction length.\cite{Zheng2024}
    
    In the MVA- and CSA-based studies it was observed that stronger laser fields lead to an increase in energy, but come at the cost of decreased beam polarization. A possible mitigation tactic is the ``dual-pulse MVA'' scheme \cite{Reichwein2022}, where two laser pulses propagate side-by-side leading to a formation of a separate ion filament that is better shielded from depolarizing field components. A problem with this scheme, however, is the difficult realization of beam alignment in experiment.

    Another approach to forming these central filaments is ion acceleration using Laguerre-Gaussian (LG) laser pulses which carry orbital angular momentum (OAM) and are characterized by an azimuthal index $\ell$ and a radial index $p$.\cite{Allen1992}

    Various theoretical studies have shown that -- due to their inherently different field structure from Gaussian pulses -- LG modes can be utilized to accelerate ion beams with a low divergence.\cite{Zhang2014, Pae2020, Hu2022, Culfa2023}

    Similarly, so-called ``light-spring pulses'' consisting of several LG modes have recently been proposed as an option for high-quality proton acceleration.\cite{Guo2024}

    Currently, the experimentally achievable intensity and maximum order of LG modes is still limited. In 2020, Wang \textit{et al.} presented results of a relativistic hollow laser with peak intensities of up to $I \approx 6.3 \times 10^{19}$ W/cm$^2$.\cite{Wang2020} With currently available and near-future laser facilities like SULF and SEL \cite{SEL}, even higher-intensity LG modes could soon be realized.

    In the context of polarized beams, Wu \textit{et al.} have already proposed the use of LG beams for electron acceleration.\cite{Wu2019} It was shown that due to the comparatively weaker azimuthal field, pre-polarized electrons maintained a higher polarization degree than with a Gaussian pulse.
    Further, they observed that the injection phase is the most crucial since polarization is only marginally affected once the particles reach a Lorentz factor of $\gamma \gg 1$.
    
    In this paper, we propose to utilize Laguerre-Gaussian laser pulses with $\ell = 1, p = 0$ to accelerate spin-polarized Helium-3. Utilizing 3D particle-in-cell (PIC) simulations, we show that the different geometry from conventional Gaussian pulses leads to a much higher degree of polarization, comparable to the dual-pulse scheme.\cite{Reichwein2022}
    In section \ref{sec:pic} we will discuss the setup and the results of our 3D-PIC simulations. These results are further discussed in section \ref{sec:discussion}.

    \section{Particle-in-cell simulations} \label{sec:pic}
    Three-dimensional particle-in-cell (PIC) simulations were conducted with the fully electromagnetic code \textsc{vlpl}.\cite{Pukhov1999, Pukhov2016} The simulation domain has a size of $x \times y \times z = 120 \times 60 \times 60 \lambda^3$ ($x$ being the direction of laser propagation; $\lambda = 800$ nm). The grid resolution is $h_x = 0.05 \lambda, h_y = h_z = 0.125 \lambda$.
    As we use the rhombi-in-plane solver \cite{Pukhov2020}, we choose the time step $\Delta t = h_x / c$. Further, we make use of \textsc{vlpl}'s scaling feature which increases the transverse grid size by a factor of $1.05$ for $|y|,|z| \geqslant 20 \lambda$ for reasons of computational efficiency.
    For the laser pulse, we only consider a Laguerre-Gaussian laser pulse with azimuthal index $\ell = 1$ and radial index $p = 0$, since realizing higher-order modes in experiment is still a significant challenge. 
    The pulse has a focal spot size of $6\lambda$ and a pulse duration of 20 fs. Its normalized laser vector potential $a_0 = e E_L / (m_e c \omega_0)$ is varied in the range $a_0 =  20-50$.

    The Helium-3 target is modeled as an initially unionized slab of $60\lambda$ length and $0.3n_\crit$ density. We consider a fully pre-polarized target, i.e. the initial spin direction for all Helium ions is the $+x$-direction.

    The spin dynamics are incorporated into \textsc{vlpl} in form of the T-BMT equation \cite{Thomas1926, Bargmann1959} which calculates the change of the semi-classical spin vector $\ssb$ as
    \begin{align}
        \frac{\D \ssb}{\D t} = - \Omegab \times \ssb \; .
    \end{align}
    The precession frequency is given as
    \begin{align}
        \Omegab = \frac{qe}{mc} \left[ \Omega_B \Bb - \Omega_v \left( \frac{\vb}{c} \cdot \Bb \right) \frac{\vb}{c} - \Omega_E \frac{\vb}{c} \times \Eb \right] \; , 
    \end{align}
    with $qe$ denoting the ion charge, $m$ its mass, and $c$ the vacuum speed of light.
    $\Eb, \Bb$ correspond to the electromagnetic field the particle is subject to, while $\vb$ is its velocity.
    The prefactors further depend on the anomalous magnetic moment $a$ and the Lorentz factor $\gamma$,
    \begin{align}
        \Omega_B = a + \frac{1}{\gamma} \; , && \Omega_v = \frac{a \gamma}{\gamma + 1} \; , && \Omega_E = a + \frac{1}{\gamma + 1} \; .
    \end{align}

    The anomalous magnetic moment for an electron is $a_e \approx 10^{-3}$. For the Helium-3 considered here, its magnitude is significantly larger, $a \approx - 4.184$.

    As discussed by Thomas \textit{et al.}, other spin-related effects like the Stern-Gerlach force or the Sokolov-Ternov effect can be neglected in our parameter regime due to the field strength and timescale we consider.\cite{Thomas2020}

    For dense targets (here: $0.3 n_\crit$), the dominant acceleration mechanism in our simulations is identified to be Magnetic Vortex Acceleration.\cite{Nakamura2010}

    In conventional MVA with a single Gaussian laser pulse \cite{Park2019}, the pulse generates a plasma channel due to its ponderomotive force. At the channel center, electrons are accelerated in the wake induced by the pulse, generating a strong forward current. Accordingly, a return current forms along the channel wall which, in turn, leads to the magnetic vortex structure. At the back edge of the target, the fields expand upon entering the vacuum region, leading to the formation of accelerating and focusing fields.

    \begin{figure}
        \centering
        \includegraphics[width=0.5\textwidth]{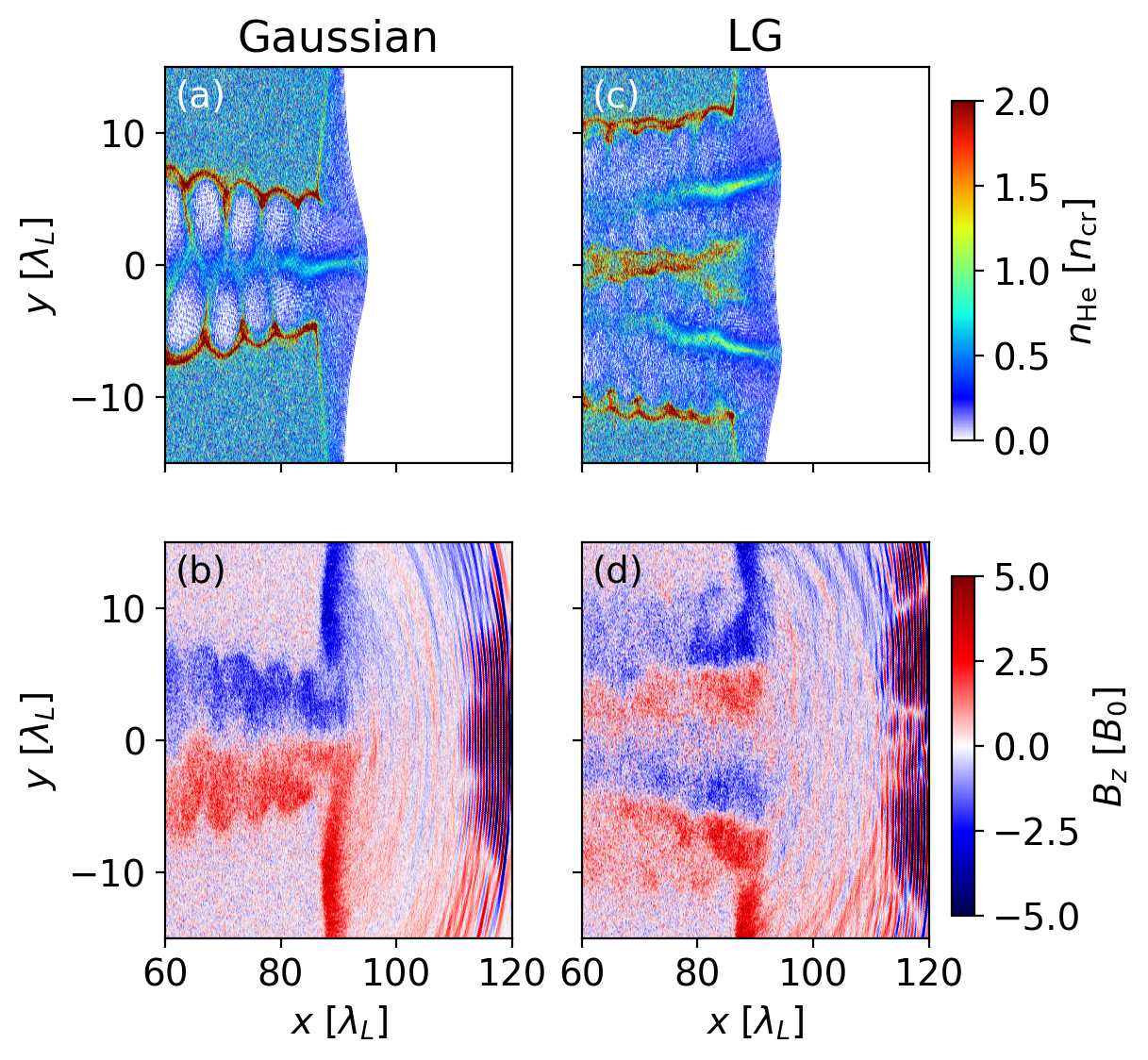}
        \caption{\label{fig:cmp_gauss}Exemplary comparison of filament formation (a), (c) and magnetic field (b), (d) for a Gaussian and an LG laser pulse.}
    \end{figure}

    By comparison, the LG pulse utilized in our simulations, creates a more complex field, which in the $x$-$y$-plane looks like two co-propagating structures (cf. Fig. \ref{fig:cmp_gauss} for a exemplary comparison of Gaussian- and LG-induced density and field structure). In fact, a similar field structure is seen in the dual-pulse MVA scheme \cite{Reichwein2022}, which essentially is nothing but a plane representation of the LG modes.
    These electromagnetic fields still lead to the formation of a well-defined filament along the optical axis $y = 0$ which is subsequently accelerated.

\begin{figure*}
        \centering
        \includegraphics[width=\textwidth]{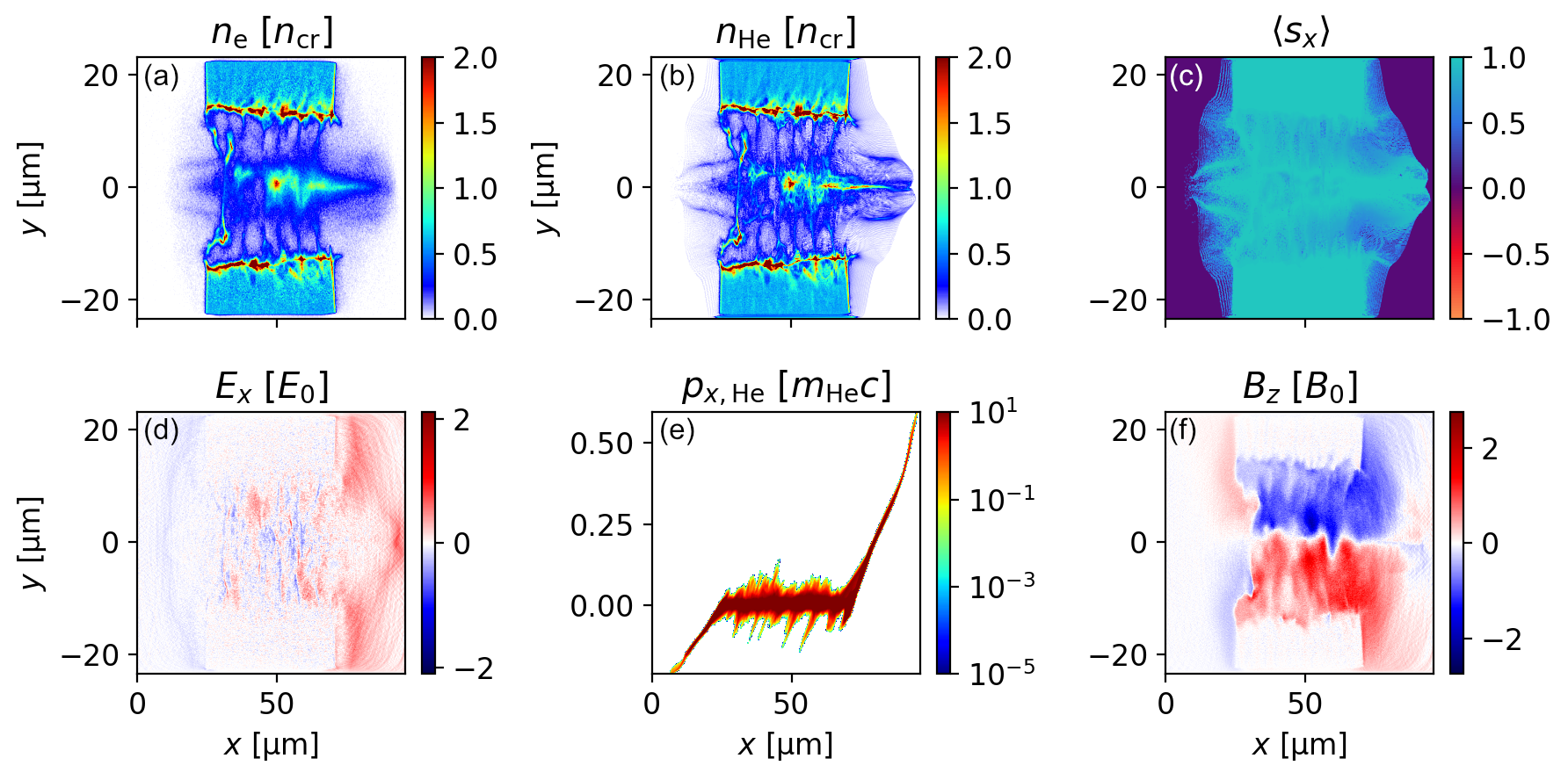}
        \caption{\label{fig:mechanism} Exemplary PIC simulation result for a short target after 470 fs. Plots (a) and (b) show the electron and Helium density, respectively. Nuclear spin polarization is shown in (c). The bottom row shows the accelerating field (d), the ions' phase space (e) and the $z$-component of the magnetic field (f).}
    \end{figure*}

    The prevalent structures in density and electromagnetic fields after approx. 470 fs is shown for the simulation with $a_0 = 50$ in Fig. \ref{fig:mechanism}. The electron and ion density [subplots (a) and (b)] exhibit the typical central filament, which -- in the case of the ions -- is highly polarized (c). The longitudinal sheath field visible in (d) accelerates the ions to high energies [cf. the phase space in (e)]. The magnetic vortex structure is indicated by the component $B_z$ in (f).

\begin{figure}
        \centering
        \includegraphics[width=\columnwidth]{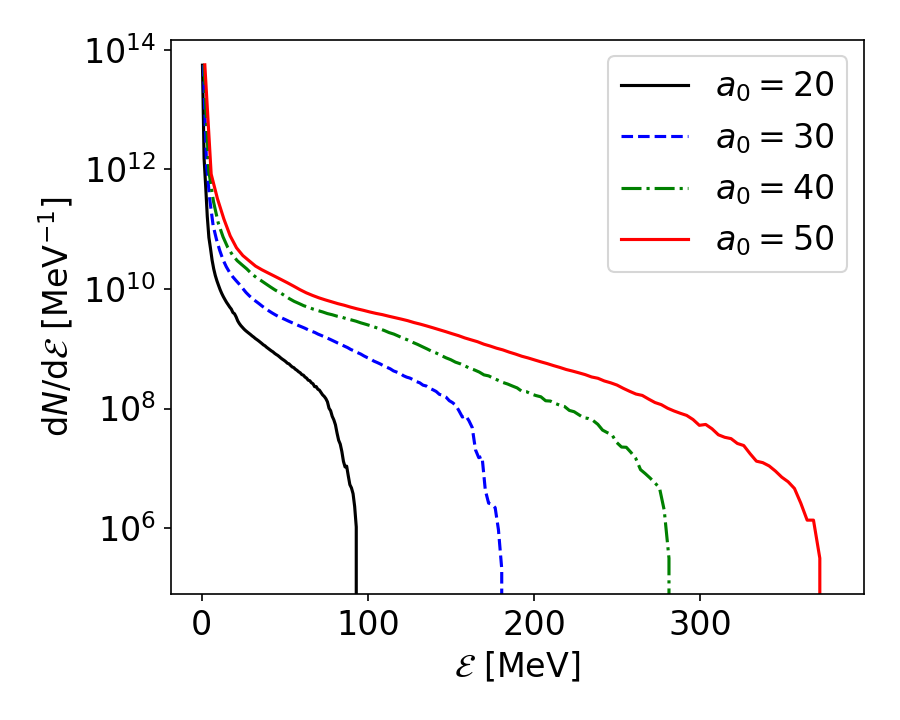}
        \caption{\label{fig:energy_short}Energy spectra for various $a_0$ and the target with $0.3n_\crit$. No distinct monoenergetic features are observed.}
    \end{figure}

    The data from the parameter scan are displayed in Tab. \ref{tab:short}.
    As expected, the maximum ion energy rises with increased $a_0$, e.g. from 91.5 MeV at $a_0 = 20$ to approx. 365.8 MeV for $a_0 = 50$. The corresponding energy spectra are shown in Fig. \ref{fig:energy_short}. No distinct mono-energetic features are observed for the laser and target parameters considered here. 

     	\begin{table}
		\centering
		
		\begin{ruledtabular}
			\begin{tabular}{lccc}
				$a_0$ & $n_e$ [$n_\crit$] & $\mathcal{E}_\mathrm{max}$ [MeV] & $P_\mathrm{min}$ [\%] \\
				\hline
				20  & 0.3 & 91.5 & 96.4 \\
				30 & 0.3 & 177.6  &  95.9 \\
				40  & 0.3 & 276.9 & 93.7 \\
				50 & 0.3 & 365.8  & 93.9 \\
                \hline
                50 & 0.03 & 35.5 & 99.3 \\
                50 & 0.006 & 4.2  & 99.9
			\end{tabular}
		\end{ruledtabular}
		
		\caption{\label{tab:short} Results from 3D-PIC simulations using a short target ($L_\mathrm{ch} \equiv 60\lambda$). Note that the minimum polarization $P_\mathrm{min}$ corresponds to the minimum polarization per energy bin (cf. Fig. \ref{fig:pol}).}
	\end{table}

\begin{figure}
        \centering
        \includegraphics[width=\columnwidth]{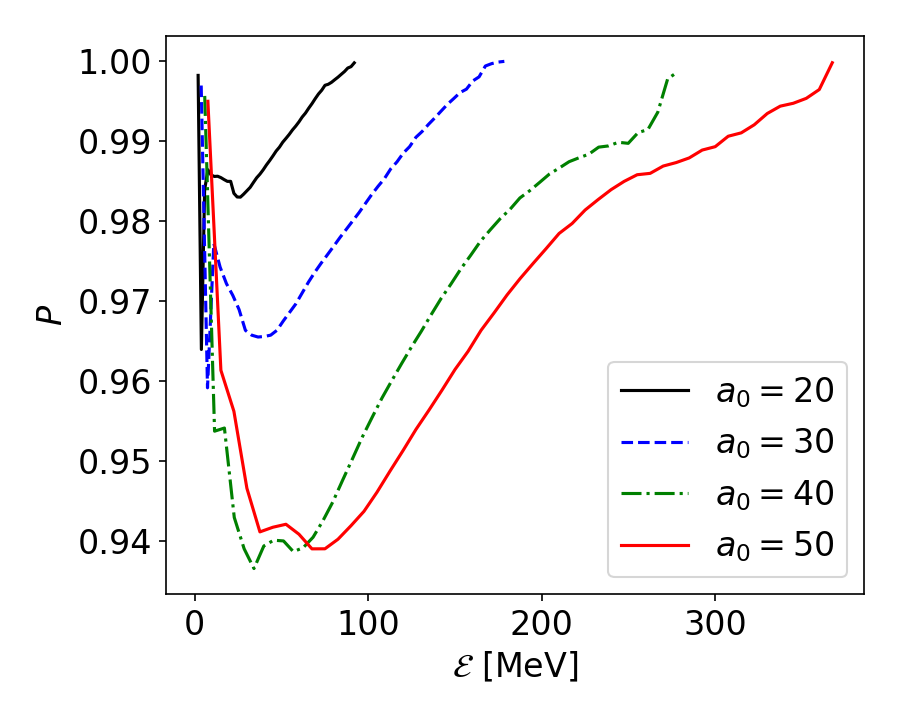}
        \caption{\label{fig:pol}Polarization spectra for the target with $0.3n_\crit$. Note that very high polarizations for large energies are partly of statistical nature, since the energy bins contain only a few macro-particles.}
    \end{figure}

    The polarization per energy bin is calculated as $P = \sqrt{P_x^2 + P_y^2 + P_z^2}$, where $P_j = \sum_i s_{i,j} / N$ is the average over the spin components of the $N$-particle ensemble in the spatial direction $j \in \lbrace x,y,z\rbrace$. The results are shown in Fig. \ref{fig:pol}. As for previous studies, the degree of polarization decreases for higher intensities since the precession frequency of spin is $|\Omegab| \propto \mathrm{max}\lbrace |\Eb|, |\Bb| \rbrace$.
    Throughout the intensity scan, polarization remains at a level $> 90\%$, which is significantly higher than what was observed in MVA with Gaussian pulses.\cite{Jin2020, Reichwein2022}
    Notably, the polarization for $a_0 = 40$ and $a_0 = 50$ is comparable, which can be attributed to the fact that the target (density) is not matched to the different intensities but is kept fixed throughout our scan.

\begin{figure}
        \centering
        \includegraphics[width=\columnwidth]{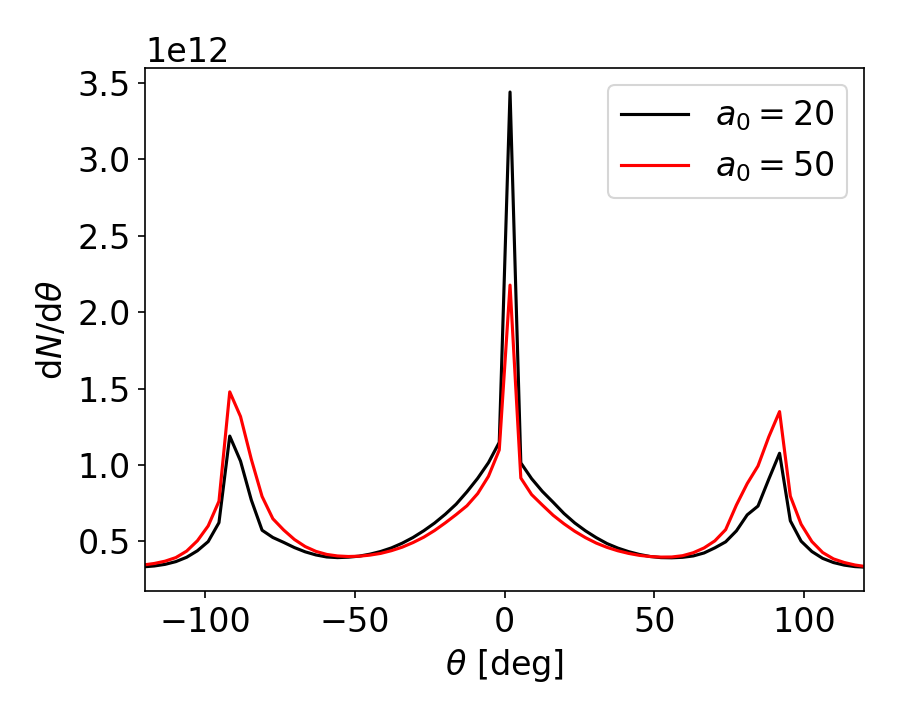}
        \caption{\label{fig:angular}Angular spectra for $a_0 = 20$ and $a_0 = 50$. The other curves are excluded for better visibility. Note that for higher intensity, fewer ions are accelerated forward.}
    \end{figure}

    The angular spectra in Fig. \ref{fig:angular} show that for the case of $a_0 = 50$ fewer ions are forward-accelerated as compared to $a_0 = 20$ which, again, is due to the absence of any target parameter matching (see also the discussion in section \ref{sec:discussion}).
    Changing the laser and target parameters further away from ``optimal MVA conditions'' would start to introduce stronger contributions of other acceleration mechanisms: e.g., in the case of the experimental study on polarized Helium-3 at PHELIX \cite{Zheng2024}, the longer, low-density target as well as the longer, weaker laser pulse led to the dominance of Coulomb explosion in $\pm 90^\circ$ rather than forward acceleration.

    The target parameters available in experiment are indeed what is currently limiting the acceleration of polarized ions via laser-plasma interaction. While high-density, short gas jets can generally be realized, the experiments with polarized Helium-3 require pre-polarized targets. These are -- up to now -- only available with a maximum density of approx. $10^{19} \mathrm{cm}^{-3} = 0.006 n_\crit$. Moreover, the target used by Zheng \textit{et al.}\cite{Zheng2024} has an interaction length on the scale of one millimeter. Simulations by Gibbon \textit{et al.} have shown that even a slightly shorter target could lead to a more forward-directed beam.\cite{Gibbon2022}

    Therefore, we have also conducted additional simulations with reduced densities but the same interaction length of $60\lambda$ to evaluate the effect on MVA via Laguerre-Gaussian pulses.
    The density reduction leads to a strongly reduced maximum energy of 35.5 MeV of $0.03 n_\crit$ and only 4.2 MeV at $0.006 n_\crit$. The minimum polarization in both cases exceeds the higher-density simulations at approx. 99\%. The higher polarization can be attributed to the improved collimation process. Whereas the central filament for the higher-density target [see Fig. \ref{fig:mechanism}(b)] has distinct kinks and imperfections, the filament at lower densities is much smoother and narrower. Thus, the particles in the filament will experience fewer depolarizing contributions.
    As mentioned before, laser and target parameters have not been matched for these simulations. We will discuss aspects of possible optimization in the following section.

    \section{Discussion} \label{sec:discussion}

    The PIC results show distinct improvements over conventional MVA with Gaussian laser pulses, and even the dual-pulse scheme.\cite{Reichwein2022} The main reason behind these improvements is the inherently different field structure of the Laguerre-Gaussian pulses:
    the potential for a Laguerre-Gaussian pulse\cite{Wilson2023} with arbitrary $\ell$ and $p = 0$ is
    \begin{align}
        a(r,\theta, x) = & a_0 \exp \left[ - \left(\frac{t}{d_0}\right)^2 - \left( \frac{r}{w_0} \right)^2 \right] \left( \frac{r}{w_0}\right)^{|\ell|} \notag \\
        & \times \exp[i(kx - \omega t - \ell \theta] \; .
    \end{align}
    Thus, for $\ell = 1, \, p=0$, the potential will have a radial dependency $a(r) \propto \exp(-r^2) r$. We will ignore the $\theta$- and $x$- dependencies for now, since we are interested in the radial force exerted by the pulse.
    The ponderomotive force is calculated as $F_p \propto - (1/\gamma) \nabla a^2(r)$, where $\gamma = \sqrt{1 + a^2(r)}$.\cite{Quesnel1998} Thus, in our case,  
    \begin{align}
        F_{p,\perp} (r) \propto \frac{4 \exp\left( -2 r^2 / w_0^2 \right) \cdot r^3 /w_0^4 - 2 \exp\left( -2 r^2 / w_0^2 \right) \cdot r /w_0^2 }{\sqrt{1  +  \exp\left( -2 r^2 / w_0^2 \right) \cdot  r^2 / w_0^2 }}
    \end{align}
    
    This force expels ions  for $r > w_0$, but focuses ions close to the central axis. Notably, the intensity of the LG mode is zero on the optical axis.

    Thus, particles in the central accelerated filament are compressed by the ponderomotive force. Spin precession is mainly induced by the prevalent azimuthal magnetic field, as observed in other MVA studies. \cite{Jin2020, Reichwein2021, Reichwein2022}
    In order to understand the high degree of polarization in the LG case better, we devise a strongly simplified model to reproduce the structure of the azimuthal magnetic field: using the ponderomotive force we can obtain the radial current $j_r$ and plug this expression into the continuity equation $\del_t \rho + \nabla \cdot \jb = 0$.
    For simplicity, we assume that the system is in steady-state, i.e. $\del_t \rho = 0$, and -- averaging out any $\theta$-dependencies -- we obtain
    \begin{align}
        \frac{1}{r} \partial_r (r j_r) = - \partial_x j_x \; ,
    \end{align}

\begin{figure}
        \centering
        \includegraphics[width=0.5\textwidth]{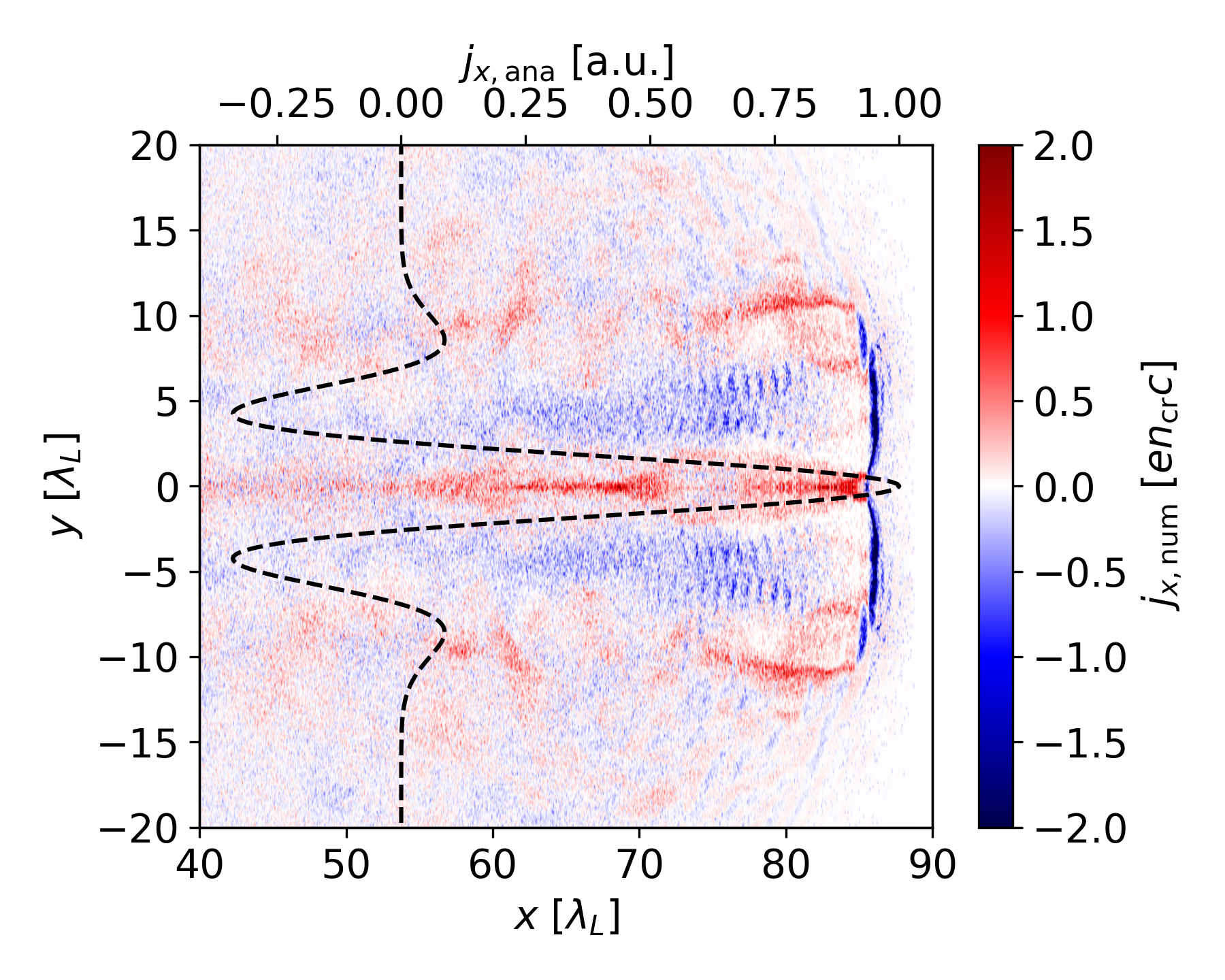}
        \caption{\label{fig:jx} The heatmap shows the longitudinal current density $j_x$ in the $x$-$y$ plane for an exemplary simulation. The black dashed line is the current density profile obtained from analytics. While the regions of forward and backward current are reproduced, the relative strength of these regions deviates from PIC simulations due to the strongly simplified nature of the model.}
    \end{figure}

    The longitudinal current profile obtained from this simplified model reproduces the regions of backward and forward currents (cf. dashed line in Fig. \ref{fig:jx}), however with strong differences in the relative magnitude of those currents compared to PIC simulations.

    Finally, the azimuthal magnetic field generated is calculated using $\nabla \times \Bb = \mu_0 \jb$, and we find
    \begin{align}
        B_\theta (r) \propto \frac{2 \exp\left( -2 r^2 / w_0^2 \right) r^3 (-2r^2 + w_0^2)}{\sqrt{1 + \exp\left( -2 r^2 / w_0^2 \right) \cdot (r/w_0)^2} w_0^4} \; . 
    \end{align}

    Again, due to the simplified nature of the model, the relative magnitude of the extrema do not coincide with PIC simulations, but the general behavior with respect to the radial coordinate $r$ does. Most importantly, the analytics reproduce the fact that there only is a weak azimuthal field component around $r=0$. Therefore, particles close to the optical axis only experience small contributions to spin precession from the azimuthal magnetic field, which ensures the high degree of beam polarization.
    The mechanism here is similar to the dual-pulse MVA scheme\cite{Reichwein2022}, where two Gaussian pulses propagating side-by-side were used. The dual-pulse scheme already presented an improvement over the single-pulse approach, generating similar field structures as in the LG case, however only in one plane.

    As we have seen in the different PIC simulations, the matching of laser and target parameters is crucial for efficient Magnetic Vortex Acceleration.
    For conditional MVA, such a condition relating laser and target parameters was found by Park \textit{et al.}:
    \begin{align}
        \frac{n_e}{n_\crit} = \sqrt{2} K \left( \frac{P}{P_\crit}\right)^{1/2} \left( \frac{c \tau}{L_\mathrm{ch}} \right)^{3/2} \; .
    \end{align}

    Here, $P_\crit$ is the critical laser power for self-focusing, $L_\mathrm{ch}$ is the interaction length, and $K = 1 / 13.5$ is a geometrical factor.
    This condition shows that increases in laser power need to be accommodated by increases in target density, if all other parameters remain fixed. Alternatively, longer targets can be employed.

    In the context of pre-polarized sources, however, the choice of target parameters is rather restrictive, as not only lower densities are required \cite{Zheng2024, Fedorets2022}, but also the interaction volume can be restricted: Sofikitis \textit{et al.} proposed an experimentally feasible target based on hydrogen halides, where the degree of polarization strongly depends on the available volume.\cite{Sofikitis2025}

    For the specific case of Laguerre-Gaussian laser pulses, the critical power ratio is modified compared to the Gaussian case \cite{Willim2023}: for modes with $\ell \neq 0, p=0$, it scales as
     $P/P_\crit = (2|\ell|)! / (4^{|\ell|}|\ell|!(|\ell|+ 1)!) P/P_G$, where $P_G$ denotes the critical power for a Gaussian laser. Thus, self-focusing of LG pulses for a given power can differ significantly and change the propagation of the laser pulse in the medium and the formation of related instabilities.

    Besides the effective interaction length for MVA, the density ramp at the back edge of the target is of importance. Nakamura \textit{et al.} first calculated that the energy per nucleon essentially scales as $\mathcal{E} / A \propto n_1 / n_2$, where $n_1$ and $n_2$ correspond to the target density before and after the ramp.\cite{Nakamura2010}
    For polarized particle beams it was seen that ramps of significant length could influence the final polarization, as ions of different polarization states were piled up.\cite{Reichwein2021} In either case, shorter ramps were shown to be beneficial for the general collimation of the ion beam since the electromagnetic fields do not have enough time to expand in the direction transverse to laser propagation. 
    From an experimental point of view, however, this aspect is limited by the available polarized ion sources as well.

    \section{Conclusions}
    In this paper we have shown the acceleration of polarized Helium-3 to several hundred MeV using Laguerre-Gaussian laser pulses ($\ell = 1, p=0$) in the range of $a_0 = $20-50. The helical structure of these modes provides beneficial electromagnetic fields for preservation of polarization as well as low-divergence beams. The minimum polarization observed is on the level of 90\% for all simulations.
    The conducted simulations clearly show that the low densities currently available from pre-polarized ion sources \cite{Fedorets2022} inherently limit their efficient acceleration: while several hundred MeV can be obtained from targets in the range of $0.3n_\crit$, realistic densities of $0.006 n_\crit$ only yield a few MeV, however with approx. 99\% polarization.
    Thus, future research on the optimization of target profiles is necessary to significantly advance the acceleration of polarized ions from laser-plasma interaction.

    \begin{acknowledgments}
        The authors gratefully acknowledge the Gauss Centre for Supercomputing e.V. \cite{GCS} for funding this project (spaf) by providing computing time through the John von Neumann Institute for Computing (NIC) on the GCS Supercomputer JUWELS at J\"ulich Supercomputing Centre (JSC). The work of M.B. has been carried out in the framework of the JuSPARC (J\"ulich Short-Pulse Particle and Radiation Center \cite{JuSPARC}) and has been supported by the ATHENA (Accelerator Technology Helmholtz Infrastructure) consortium. The work of A.P. has been supported in parts by BMBF project no. 05P24PF1 (Germany). The work of T.-P.Y. has been supported by the National Natural Science Foundation of China (Grant No. 12375244) and the Natural Science Foundation of Hunan Province of China (Grant No. 2025JJ30002).
        L.R. would like to thank Oliver Mathiak for fruitful discussions.
	\end{acknowledgments}

\bibliography{bibliography}

\end{document}